\documentstyle[preprint,pra,aps,psfig,epsfig]{revtex}
\input{psfig.sty}
\begin{document}
\draft
\title{ Dynamics of a trapped 2D  Bose-Einstein condensate
with periodically and randomly varying
atomic scattering length}
%\author{F.Kh. Abdullaev}
%\address{Physical-Technical Institute of the Uzbek Academy of
%Sciences, 700084, Tashkent-84, G.Mavlyanov str.,2-b, Uzbekistan}
%\author{J. C. Bronski}
%\address{Department of Mathematics, University of Illinois at
%Urbana-Champaign, 1409, West Green St. IL, 61801, USA}
%\author{R.M. Galimzyanov}
%\address{Physical-Technical Institute of the Uzbek Academy of Sciences, 700084,
%Tashkent-84, G.Mavlyanov str., 2-b,  Uzbekistan }
\author{F.Kh. Abdullaev$^{a}$\footnote{Corresponding author.
Fax: +7-712-35-42-91\\
{\it{E-mail address:}} fatkh@physic.uzsci.net (F.Kh. Abdullaev)
},
 J. C. Bronski$^{b}$, R.M. Galimzyanov$^{a}$}

\address{$^{a}$ Physical-Technical Institute of the Uzbek Academy of
Sciences, 700084, Tashkent-84, G.Mavlyanov str.,2-b, Uzbekistan}

\address{$^{b}$ Department of Mathematics, University of Illinois at
Urbana-Champaign, 1409, West Green St. IL, 61801, USA}

\maketitle
\begin{abstract}

In this work we consider the oscillations and associated resonance
of a 2D Bose-Einstein condensate under periodic and random
modulations of the atomic scattering length.
For random oscillations  of the trap potential and of the atomic
scattering length we are able to calculate the mean growth rate
for the width of the condensate. The results obtained from the reduced
ODE's for oscillations of the width of condensate are
compared with the numerical simulations of the full 2D
Gross-Pitaevskii equation with modulated in time coefficients.

\end{abstract}

\newpage
\section{Introduction}

The problem of the oscillations of two and three  dimensional
trapped Bose-Einstein condensates (BEC) have attracted
a great deal of  recent attention\cite{Alfa}, particularly
the problem where the trapping potential is temporally modulated.
The main physical motivation for considering
such a problem is the experiments of Jin, et. al.\cite{Jin},
which observed resonant response of the  BEC oscillations
to periodically perturbed traps.
Some theoretical
investigations of the problem have included
the work of Castin and Dum\cite{Castin}, in which the resonances
in the width oscillations have been analyzed using scaling theory;
the work of Garcia\cite{Garcia}, in which it was argued
that the oscillations of 2D condensate can be described by the parametric
resonances of linear oscillator; and the papers by Pitaevski and
Rosch\cite{Pit1,Pitaevsky} and Kagan et. al.\cite{Kag1}, where it was shown that
a 2D condensate exhibits a harmonic mode with frequency $2\omega_0$
when the harmonic trap is driven with frequency  $\omega_0$.
Note that the related  problem of a soliton interacting with an impurity
in 2D molecular crystals had been studied earlier by Gadidei et al.\cite{Gaididei}.

The excitation of resonant oscillations is also interesting
as a possible mechanism for the stabilization of a BEC with
attractive interactions. Such a possibility has been demonstrated
for the BEC  with  negative atomic scattering length under temporally
modulated trap potential if the  surface mode is exited \cite{Pat}. In this
work a variational approach is used, and the trap was anisotropic.

It is also interesting to consider the response of a condensate
to temporally periodic variations in the atomic scattering length.
Such a  variation of the atomic scattering length can be achieved
experimentally by varying the magnetic field \cite{Inouye} or using
optically induced Feshbach resonances \cite{Fatami}.
The case where the scattering length varies  monotonically in
time, in particular the case where the  sign changes from
repulsive to attractive, has been studied recently by Dalfovo et. al\cite{Alfa},
and Fedichev, et. al.\cite{Kagan}. Similarly the effect of a periodic
variation of atomic scattering length on the tunneling between two condensates
in the double-well trap, a resonant tunneling, has been studied
by Abdullaev and Krankel\cite{Abd1}.

These observations suggest the possibility of new phenomenon in
Bose-Einstein condensate width oscillations when the atomic scattering
length $a_s$ is allowed to fluctuate about some mean value.
In this paper we study the influence of both periodic
and random fluctuations of the atomic scattering length on the condensate
dynamics.
As noted in [6], at temperatures $T \gg ng, g= 4\pi \hbar^2 a_{s}/m$,
(where $n$ is the gas density, $a_{s}$ is the atomic scattering length, $m$
is the atom mass), only the condensate evolution is pronounced. The
perturbation of the thermal cloud is small at these temperatures, and
in this case the eigenfrequencies of small oscillations of the thermal
cloud are close to those of the condensate.

The specific problem that we consider is the 2D Gross-Pitaevskii equation
with periodic or random variations in the coefficient of the nonlinear term.
This problem is interesting not only for
BEC but also for nonlinear optics too -  for example the dynamics of
optical beams in nonlinear layered waveguides
\cite{BergeOL}.

The structure of the remainder of the article is follows: In Sect.2
we describe the model and derive the reduced ODE system to describe
the dynamics of the condensate width. In Sect.3 we analyze the
effect of time-period perturbations on the width dynamics
using the action-angle variables for the reduced ODEs. We
explore the resonances in condensate oscillations in Sect.4,
and the dynamics of width oscillations under fluctuating trap
potential and atomic scattering length in last section.

\section{The Reduced ODE Model}

It is well-known that wavefunction for a 2D Bose-Einstein condensate in
a trap potential $V(r)$ is described by the
Gross-Pitaevskii equation:
\begin{equation}\label{NLSE}
i\hbar \psi_t = -\frac{\hbar^2}{2m}\Delta \psi + V_{tr}(r)\psi +
g(t) |\psi|^2 \psi.
\end{equation}
Here $V_{tr}(r) = m\omega^2 r^2 /2$ is the trap potential and
$g(t) = 4\pi \hbar^2 a_s/m$, with $a_s$ being the atomic scattering length.
We will assume that the time-dependent scattering length is
constant to leading order, with
$$a_s = a_0 (1 + \epsilon_{0}(t)).$$ In this paper we analyze the
case where $\epsilon_{0}(t)$, representing the small fluctuations,
is a  periodic function of time as well as the case
where $\epsilon_{0}(t)$ is a  mean zero white noise random process:
$<\epsilon_{0}(t)\epsilon_{0}(t')> = 2\sigma^2 \delta(t-t')$.

Under the scaling
$\tau = t\omega$,$x' = x \sqrt{\hbar/(2m\omega)}$,
$u = \psi \sqrt{4 \pi a_0/\omega}$
the Gross-Pitaevskii equation can be written in the dimensionless form
\begin{equation}\label{nlse}
iu_{\tau}  + \Delta u - \frac{r^2}{2}u + s|u|^2 u = 0,
\end{equation}
where $s >0, s<0 $ correspond to attractive or repulsive
interactions between atoms in the condensate respectively.

A number of methods can be employed to investigate the dynamics of the
Bose-Einstein condensate under temporal variations of
the atomic scattering length. One of the simplest to apply is the averaged
Lagrangian approach.
According to this method we take the Gaussian anzatz for the field
 \cite{Anderson,Stoof,Garcia}.
\begin{equation}\label{anz}
u(r,\tau) = A(\tau) \exp(-\frac{r^2}{2a(\tau)^2} + \frac{ib(\tau)r^2}{2}
 + i\phi(\tau)),
\end{equation}
To derive the equations for the wavepacket parameters
$A(\tau),a(\tau),b(\tau),\phi(\tau)$
one should calculate the averaged Lagrangian
$$\bar{L}(\tau) = \int r dr L(r,\tau).$$ For the Gaussian ansatz in
Eq.(\ref{anz}) the averaged Lagrangian is given by
\begin{equation}
%\bar{L} = -\frac{\pi}{2}A^{2}a^{2}(a^{2}b_{\tau} + \phi_{\tau} + \frac{2}{a^2} +
\bar{L} = -\frac{\pi}{2}A^{2}a^{2}(a^{2}b_{\tau} +2\phi_{\tau} + \frac{2}{a^2} +
2a^{2}b^2 +
a^2 - \frac{s(\tau) A^2}{2} ).
\end{equation}

The Euler-Lagrange equations for the functional $\bar L(\tau)$
lead to the following equations for the
the width $a(\tau)$ and chirp $b(\tau)$.
\begin{eqnarray}
a_{\tau} = 2ab \nonumber\\
b_{\tau} = \frac{2 - s(\tau) N}{a^4} - 2b^2 - 1 ,
\end{eqnarray}
with $N = \frac{1}{2\pi}\int |u|^2 d^{2}r = \frac{a^{2}A^{2}}{2}.$
It is easy to see that the above is a Hamiltonian system, with the
width $a$ playing the role of a position variable, and the chirp $b$
the conjugate momentum. One can eliminate $b$ from the above system
which leads to the following evolution equation for $a$:

\begin{equation}\label{ode}
a_{tt} + a = \frac{2 - s(t) N}{a^3},
\end{equation}

where $t = \sqrt{2}\tau$.
For the attractive problem with constant atomic scattering length
$s(t) = 1$  the variational approach predicts the  critical
threshold for collapse $N_c = 2$, compared with the the exact
value $N_c = 1.862$.

We comment that, if the initial condition is close to that of the ground
state (Townes) soliton, $|N - N_{c}| \sim \epsilon \ll 1$,
then the modulation theory of Fibich and Papanicolaou\cite{Fibich} leads
naturally to a modulation equation of the same form as the above, where
again the constant $2$ is replaced by the critical value $ 1.862$.
The qualitative types of behavior exhibited by the above equation do not,
of course, depend on the exact values of the constants, since all such
constants can be scaled out.

It is useful to rewrite Eq.(\ref{ode})  in the form
\begin{equation}\label{posc}
a_{tt} + a = \frac{Q(t)}{a^3},
\end{equation}
where $Q(t) = Q (1 + \epsilon(t)), Q = 2 - N, \epsilon =
N\epsilon_{0}/(2-N)$.
In the next section we construct the action-angle variables for the
above problem.

\section{Perturbation theory in the action-angle variables}

We begin our analysis of the Eq.(\ref{ode}) by constructing the
action-angle variables. Of course any two dimensional
Hamiltonian system can always be reduced to quadrature, but the above
system is particularly nice because the action-angle variables
can be expressed in terms of elementary functions.
First we note that Eq.(\ref{ode}) is Hamiltonian:
\begin{equation}
%H(a_t,a) = \frac{a_t^2}{2} + U(a) = \frac{a^2}{2} + \frac{Q}{2a^{2}}.
H(a_t,a) = \frac{a_t^2}{2} + U(a), U(a) = \frac{a^2}{2} + \frac{Q}{2a^{2}}.
\end{equation}

For $N < 2, Q > 0$ and the motion of the effective particle is bounded, and
bounded away from $0$ due to the ``angular momentum'' barrier
$\frac{Q}{2a^{2}}$.
If the energy is $E$ then the width oscillates between
$a_{min}=\sqrt{E-\sqrt{E^2-Q}}$ and $a_{max}=\sqrt{E+\sqrt{E^2-Q}}$.
The minimum of the potential $U$ occurs at $a_{c} = Q^{1/4}$, with a
minimum energy  of  $E=U_{c}= \sqrt{Q}$.
When $N > 2 , Q < 0$ there is no local minimum, and there exist solutions
for which $a \rightarrow 0$, corresponding to collapse of a condensate
(see Fig.\ref{fig_1}).

%%%%%%%%%%%%%%  Fig.1     BEGIN    %%%%%%%%%%%%%%%%%%%%%%%%%%%%%%%
%\begin{center}
%\mbox{\hspace{-0.35in}
%\psfig{figure=fig_1.eps,height=3.8in,angle=-90.}}
%\end{center}
%  \vskip 0.2in
%{\protect\small
%   FIG.  \ref{fig_1}.
%}
%\vskip0.2in
%
%\noindent
%%%%%%%%%%%%%%  Fig.1     END  %%%%%%%%%%%%%%%%%%%%%%%%%%%%%%%

Since the Hamiltonian is conserved, $H = E$, we have
\[
\int \frac{da}{\sqrt{2 E - a^2 - Q a^{-2}}} = \int dt
\]
which can be integrated up to the following solution for $Q > 0:$
\begin{equation}
a(t) = \sqrt{E + \sqrt{E^2 - Q}\sin(2t + \psi_0)}.\label{eqn:soln}
\end{equation}

The action variable for a Hamiltonian system is defined to be
\begin{equation}
J = \frac{1}{2\pi}\oint p dq,
\end{equation}
where $q$ is the position variable and $p$ the conjugate momentum.
For the above oscillator the position variable is $a$, with momentum
$p = \sqrt{2Ea^2 - a^4 -Q}/a.$
The integration is easily done via contour integration in the complex
$a$ plane leading to the following expression for the action vaiable:
\begin{equation}
J = \frac{1}{2}(E - \sqrt{Q}).
\end{equation}
The bottom of the potential well, $E = \sqrt{Q}$, corresponds to $J = 0$.
Since the energy is linear in the action, $E = 2 J + \sqrt{Q},$
the frequency of the unperturbed oscillations is constant
\begin{equation}
\omega = \frac{dE}{dJ} = 2.
\end{equation}
This can, of course, also be seen directly from the solution given in
Eq. (\ref{eqn:soln}).

The Hamiltonian for the perturbed problem is given by
$$H =p^2/2 + U(a) +\epsilon(t) V =
p^2/2 + U(a) + \frac{Q \epsilon(t)}{2 a^2}$$,
with $V(a)$ the perturbation Hamiltonian. In the
action-angle coordinates $J,\theta$ the perturbation becomes
$$V_{\theta} = -\frac{Q\sqrt{E^2 - Q}\cos(\theta)}
{2(E + \sqrt{E^2 - Q}\sin(\theta))^2}.$$
and the perturbed equations of motion are given by
\begin{eqnarray}
\frac{dJ}{dt} = -\epsilon(t)\frac{\partial V}{\partial \theta},
\label{eqn:ode1}\\
\frac{d\theta}{dt} = 2 + \epsilon(t)\frac{\partial V}{\partial J}.
\label{eqn:ode2}
\end{eqnarray}

\section{Resonances of BEC oscillations under periodically varying atomic
scattering length}
In this section we consider perturbations $\epsilon(t)$ which are periodic
in time. For simplicity we discuss the case $\epsilon(t) = \sin(\Omega t)$,
though this can easily be generalized.
To analyze the resonances of a BEC under such perturbations
we use the multiscale expansion method.
We introduce the slow time $T = \epsilon t$ and assume a multiple-scales
ansatz of the form
\begin{eqnarray}
\theta = \theta^{(0)}(t, T) + \epsilon \theta^{(1)}
(t, T) + ...,\\
J = J^{(0)}(t, T) + \epsilon J^{(1)} + \epsilon J^{(1)}(t, T) + ...
\end{eqnarray}
The solution to the evolution on the fast scale is obviously
$\theta^{(0)} = 2t + \Phi, J={\rm constant}$. At the next order we find the
following evolutions for the action $J$ and the slow angle $\Phi$
on the slow scale:
\begin{eqnarray}
\frac{dJ}{dT} = -\frac{\partial \bar{H}}{\partial \Phi}\\
\frac{d\Phi}{dT} = \frac{\partial \bar{H}}{\partial J},
\end{eqnarray}
where $\bar H$, the slow Hamiltonian, is given by
 $$\bar{H}(J,\Phi) = \frac{1}{2\pi}\int_0^{\pi}\frac{\sin(\Omega t)dt}
{(2J + Q^{1/2} + 2J^{1/2}(J + Q^{1/2})^{1/2}\sin(2 t + \Phi))}.$$

It is clear that we have a resonant response whenever $\omega = 2 n$,
and the period of the perturbation is commensurate with the
natural period of the condensate. Note that, even though this is a
nonlinear oscillator the period is {\em independent} of the amplitude.
The effect of this is that there is no "detuning" from the resonant
frequency as the amplitude increases. Of course when $J$ becomes
large and the width oscillations become significant we no longer
expect the averaged Lagrangian ODE's to provide an accurate description
of the dynamics of the condensate width.

The first (nontrivial) resonance occurs for $\Omega = 2.$
In this case the slow Hamiltonian is given by
\begin{equation}
\bar{H}(J,\Phi) = -\frac{\cos(\Phi)J^{1/2}}{2(J + Q^{1/2})^{1/2}Q^{1/2}}.
\end{equation}
and the equations of motion are
\begin{eqnarray}
\frac{dJ}{dt} &=& -\frac{\partial \bar{H}}{\partial\Phi} =
-\frac{\sin(\Phi)J^{1/2}}{2(J + Q^{1/2})^{1/2}Q^{1/2}}, \\
\frac{d\Phi}{dt} &=& \frac{\partial \bar{H}}{\partial J} =
-\frac{\cos(\Phi)}{4 J^{1/2}(J+Q^{1/2})^{3/2}}.
\end{eqnarray}
The phase plane for the oscillator is depicted in Fig. (\ref{fig_2}).

Note that the line $\Phi = -\pi/2$ is invariant under the dynamics, and
along this line $J$ evolves according to
\begin{equation}
\frac{dJ}{dt} =\frac{J^{1/2}}{2(J + Q^{1/2})^{1/2}Q^{1/2}}.
\end{equation}
This solution corresponds to a resonant driving, where the
variations in the scattering length reinforce the width oscillations.
It is clear that for large $t$ the action grows linearly,
$J \propto t/2Q^{1/2}$
and thus amplitude of oscillations grows like $a \sim \sqrt{t}$.
It is also clear from the phase portrait that all orbits for
which $\Phi \neq \pi/2$ are asymptotic to the invariant manifold
$\Phi = -\pi/2$, so for generic initial conditions one expects
that the width will grow like  $a \sim \sqrt{t}.$
There is, of course, also the solution $\Phi=\pi/2$ in which the variations
in the scattering length are anti-resonant with the
variations in the width of the condensate, and act to {\em damp}
the width oscillations. It is easy to see that when $\Phi = \pi/2$
the action $J$ goes to zero in finite time. Since generic initial
conditions are asymptotic to $\Phi=-\pi/2$ the $\Phi = \pi/2$ solutions
are unlikely to be observed experimentally, though it is possible that
they could be realized with some kind of control.

%%%%%%%%%%%%%%  Fig.2     BEGIN    %%%%%%%%%%%%%%%%%%%%%%%%%%%%%%%
%\begin{center}
%\mbox{\hspace{-0.35in}
%\psfig{figure=fig_2.eps,height=3.8in,angle=-90.}}
%\end{center}
%  \vskip 0.2in
%{\protect\small
%   FIG.  \ref{fig_2}
%}
%\vskip0.2in
%
%\noindent
%%%%%%%%%%%%%%%  Fig.2     END    %%%%%%%%%%%%%%%%%%%%%%%%%%%%%%%
Next resonance occurs at $\Omega = 4$. In this case the Hamiltonian is
given by
\begin{equation}
\bar{H} = \frac{J \sin(2\Phi)}{(J + Q^{1/2})Q^{1/2}}.
\end{equation}
Again it is easy to see that the line $\Phi = \pi/2$ is invariant
under the dynamics, corresponding to width oscillations which are in
phase with the variation of the scattering length.
Since the action evolves according to
\begin{equation}
\frac{dJ}{dT} = -\frac{2J\cos(2\Phi)}{(J + Q^{1/2})Q^{1/2}}
\end{equation}
when $\Phi = \pi/2$ we have that the growth of the action $J$ is
asymptotically linear - $J \approx  2t/Q^{1/2}$.

\subsection{Numerical simulations}

We have conducted some numerical simulations to test the validity
of the calculations of the last section. We discretize the problem
in the standard way, with time step $\Delta t$ and spatial step $h$,
so the $u_j^k$ approximates $u(jh,k\Delta t)$. More specifically
we approximate Eq.(\ref{nlse}) with the following second order accuracte
semi-implicit Crank-Nicholson scheme,
\begin{eqnarray}\label{FDA}
\frac{i(u^{k+1}_{j} - u^{k}_{j})}{\Delta t} = -\frac{1}{2h^{2}}[
(u^{k+1}_{j-1} - 2u^{k+1}_{j} + u^{k+1}_{j+1}) +
(u^{k}_{j-1} - 2u^{k}_{j} + u^{k}_{j+1})] - \nonumber \\
\frac{1}{4 r_j h}
[(u^{k+1}_{j+1} - u^{k+1}_{j-1}) + (u^{k}_{j+1} - u^{k}_{j-1})] +
\frac{1}{2}[\frac{{r_j}^2}{2} - (1 + \epsilon(t)) s|u^{k}_{j}|^2]
(u^{k}_{j} + u^{k+1}_{j}),
\end{eqnarray}
where $\epsilon(t)$ is the perturbation term. In the numerical
simulations, as in the analysis, the perturbation was chosen to be
$\epsilon(t) = \epsilon \sin(\Omega t).$
Eq. (\ref{FDA}) represents a tridiagonal
set of equations for unknowns $u^{k+1}_{j-1}$, $u^{k+1}_{j}$
and $u^{k+1}_{j+1}$ $[j = 1,2 ... (N - 1)]$ in a lattice  of $N$ points,
with the values of $u_0^{k+1}$ and $u_{N+1}^{k+1}$ being determined
from the boundary
conditions $\frac{\partial{u}}{\partial{r}}\mid_{r=0} = 0$ and
$u(r)\mid_{r=\infty} \rightarrow 0$.

The set of algebraic equations (\ref{FDA}) is solved by the
vectorial sweep method. In actual calculations the typical space step
$h$ ranged from 0.01 to 0.005 and time step $\Delta t$ from 0.005
to 0.002 depending on the closeness of $\Omega$ to the points of
resonance.

The first experiment, shown in Fig. \ref{fig_3},
depicts the solution of the
Gross-Pitaevski equation when the driving frequency is
slightly off resonance: the natural frequency of the width
oscillations is $\Omega_0 = 2$, while the perturbation has
frequency $\Omega = 1.9$ and amplitude $\epsilon = 0.1$.
In this plot, as well as all subsequent plots, the solid line
represents the solution to the full PDE while the dotted
line represents the solution to the reduced ODEs.
We observe oscillations with
beats, representing the superposition of low and high frequency oscillations.
From simulations we seen the good agreement between the
full PDE  and the reduced ODE model.

%%%%%%%%%%%%%%  Fig.3     BEGIN    %%%%%%%%%%%%%%%%%%%%%%%%%%%%%%%
%\begin{center}
%\mbox{\hspace{-0.35in}
%\psfig{figure=fig_3.eps,height=3.8in,angle=-90.}}
%\end{center}
%  \vskip 0.2in
%{\protect\small
%   FIG.  \ref{fig_3}.
%}
%\vskip0.2in
%
%\noindent
%
%%%%%%%%%%%%%%  Fig.3     END  %%%%%%%%%%%%%%%%%%%%%%%%%%%%%%%

Fig.\ref{fig_4} represents  numerical simulations of the ODE and PDE
models at the  resonance point $\Omega =
\Omega_0 = 2$. The agreement is quite good for the period of
the width oscillations, although there is clearly some discrepancy
in the actual value of oscillation amplitude. Note that the same
phenomenon has been observed in numerical simulations of the
resonances in condensate oscillations under periodically varying
trap potential \cite{Garcia}.  The graph in Fig.\ref{fig_5} depicts the energy
versus time for the ODE (dotted line) and PDE (solid line) simulations for the same values of
parameters as in Fig.\ref{fig_4}. The agreement between simulations of the
full PDE and ODE is very good for time less than 30 or so,
though for times between $30$ and $40$ the oscillations in the
energy of the PDE are smaller than the analogous oscillations
of the ODE, probably due to radiative damping. As was argued earlier
the variational approach is
unlikely to be valid when the oscillations have large amplitude
and other effects, such as radiative damping, become important.

%%%%%%%%%%%%%%  Fig.4, 5     BEGIN  %%%%%%%%%%%%%%%%%%%%%%%%%%%%%%%
%\begin{center}
%\mbox{\hspace{-0.35in}
%\psfig{figure=fig_4.eps,height=3.8in,angle=-90.}}
%\end{center}
%  \vskip 0.2in
%{\protect\small
%   FIG.  \ref{fig_4}.}
%\vskip0.2in
%
%\noindent
%
%\begin{center}
%\mbox{\hspace{-0.35in}
%\psfig{figure=fig_5.eps,height=3.8in,angle=-90.}}
%\end{center}
%  \vskip 0.2in
%{\protect\small
%   FIG.  \ref{fig_5}.}
%\vskip0.2in
%
%\noindent
%%%%%%%%%%%%%%%  Fig.4, 5     END  %%%%%%%%%%%%%%%%%%%%%%%%%%%%%%%

Fig.\ref{fig_6} depicts the oscillations of the square width of the
condensate at the $2:1$ resonance $\Omega = 2\Omega_0$, where the
frequency of the perturbation is twice the natural period of the
width oscillation. As in
previous case the frequencies agree very well, but the
amplitude of oscillations is larger for PDE in comparison with
ODE. This discrepancy grows with time.
In this case it seems clear that the Gaussian anzatz does not
correctly capture the behavior of the underlying PDE. The rate of
growth of the energy, which is not depicted, is
still linear, in agreement with the theoretical estimates.

%%%%%%%%%%%%%%  Fig.6     BEGIN  %%%%%%%%%%%%%%%%%%%%%%%%%%%%%%%
%\begin{center}
%\mbox{\hspace{-0.35in}
%\psfig{figure=fig_6.eps,height=3.8in,angle=-90.}}
%\end{center}
%  \vskip 0.2in
%{\protect\small
%   FIG.  \ref{fig_6}.}
%\vskip0.2in
%
%\noindent
%%%%%%%%%%%%%%  Fig.6     END  %%%%%%%%%%%%%%%%%%%%%%%%%%%%%%%

\section{Evolution of BEC under random fluctuations}
The action-angle formulation also provides a suitable framework
for the analysis of condensate oscillations
under random (white-noise) perturbations. One interesting quantity which
can be calculated is the mean time to achieve a given distortion
under a stochastic perturbation. From the ODE point of view this problem
is equivalent to the
problem of the mean time to achieve the given level of the oscillations
amplitude
in the effective potential.

The starting point for this calculation is
Eqn(\ref{eqn:ode1},\ref{eqn:ode2}), where $\epsilon(t)$
is taken to be a white noise process $\epsilon(t) dt = dB$, with
$dB$ the increment of a Brownian motion $B$. This pair of
stochastic ordinary differential equations has an associated Fokker-Planck
equation for $P(J,\theta,t)$, the probability of having values
$J,\theta$ at time $t$. In the weak noise limit one can do a straightforward
multiple scale expansion on this Fokker-Planck equation.
Upon doing so and averaging over the fast angle variable
one is lead to the following equation for $P(J,t)$:
\begin{equation}
\frac{\partial P}{\partial t} = \sigma^2 \frac{\partial }{\partial J}(A(J)
\frac{\partial P}{\partial J}) ~~~~~~~P(J,0) = \delta(J-J_0),
\end{equation}
where $A(J)$ the average diffusivity for diffusion across energy levels
is given by $$A(J)=\frac{1}{2\pi}\int_{0}^{2\pi}V_{\theta}^2 d\theta.\label{eqn:diff}$$
For details of this calculation see the paper of
Abdullaev, et. al\cite{Abd} This one-dimensional diffusion can
be analyzed in some detail. For instance the mean time to
reach action $J$ staring from $J_0$ is given by
\begin{equation}\label{exit}
<\!\!t\!\!>_{J_0,J} = \int_{J_0}^{J}\frac{J dJ}{A(J)}.
\end{equation}
Note that if $A(J)$ grows sufficiently rapidly, so that the above
integral converges, then the mean time to random walk to $J=\infty$
is actually finite.

\subsection{Fluctuating trap potential}
One physically interesting random perturbation is when the
strength of the trap is allowed to fluctuate randomly.
Since the trapping potential is imposed optically, by a
laser,it is important  to take into account the fluctuations of
the effective trap potential due to the fluctuations of the laser field
intensity. The fluctuations of the laser field intensity lead to
random variations of the frequency of the effective harmonic
trap $\omega^2 = \omega_{0}^{2}(1 + \epsilon(t))$, where
$\epsilon(t)$ is the white noise process \cite{Savard}. The
function $\epsilon(t)$ should be thought of as the fluctuations
of the laser intensity around its mean value $E_0$:
$$\epsilon(t) = \frac{E(t)-E_0}{E_0}$$
This problem was studied in \cite{Abd3} using a moment expansion, though
without  numerical simulations of the full stochastic GP equation.
Here we consider this problem both as an illustration of effectiveness of
our technique, and to provide numerics for this problem. In this case the perturbation Hamiltonian is $$V(a) = \frac{1}{2}(E +
\sqrt{E^2 -Q}\sin{\theta} ).$$ Then from Eqn.(\ref{eqn:diff})
the diffusivity is given by
$$A(J) =\frac{J(J+\sqrt{Q})}{2}.$$

Calculating the mean exit time $<\!t\!>$ we find the energy as a function
of the mean exit time
\begin{equation}
E = 2\sqrt{Q}(e^{\sigma^2 <\!t\!>/2} -\frac{1}{2}).
\end{equation}

While the mean exit time is ${\em not}$ the same as the
physical time in the limit of weak noise we expect that, to
leading order in the noise parameter, the mean energy should
have the same dependence on $t$.
In Fig.\ref{fig_7} we present the results of comparison
of the theory
and the numerical simulations of the full stochastic NLS equation (\ref{nlse})
%for $\sigma = 0.04, N = 0.8;1.0$ respectively.For PDE we perform averaging
for $\sigma = 0.04, N = 1.0$. For the PDE we have averaged
over 50 realizations, and for ODE we have averaged over 1000 realizations.
The figure shows good agreement between predictions of the theory
and numerical simulations. For large times the solution to the
ODE overestimates the actual energy. This discrepancy is probably
due to effects such as dispersive radiation, excitation of higher
modes, etc. which are difficult to incorporate into the variational ansatz.

From the above we can easily estimate the mean exit time from the
potential.
In optical traps the depth of the effective potential is $U_0 = 2E_R
=2\hbar^2 k^2/2m$, with $k=2\pi/\lambda$, where $\lambda$ is the wavelength,
The condensate lifetime can be estimated as the time when become
$E \sim U_0$. Then $$t_{BEC} \approx \frac{2}{\sigma^2}
\ln(\frac{U_0}{2\sqrt{Q}} + \frac{1}{2}).$$

Substituting the typical values for the laser intensity fluctuations we find
that the mean exit time is of the order of seconds,
$t_{BEC} \sim sec$.
%%%%%%%%%%%%%%  Fig.7     BEGIN    %%%%%%%%%%%%%%%%%%%%%%%%%%%%%%%
%\begin{center}
%\mbox{\hspace{-0.35in}
%\psfig{figure=fig_7.eps,height=3.8in,angle=-90.}}
%\end{center}
%  \vskip 0.2in
%{\protect\small
%   FIG.  \ref{fig_7}.
%}
%\vskip0.2in
%
%%%%%%%%%%%%%%%  Fig.7     END  %%%%%%%%%%%%%%%%%%%%%%%%%%%%%%%

\subsection{Fluctuating atomic scattering length}
In the case where the atomic scattering length is allowed to
fluctuate we find that the perturbation Hamiltonian is
given by $$V_{\theta} = -\frac{Q \sqrt{E^2 - Q}\cos(\theta)}
{2(E + \sqrt{E^2 - Q}\sin(\theta))^2}.$$
From this is follows that the the effective diffusivity is
\begin{equation}
A(J) = \frac{Q^{2}(E^2 -Q)}{8\pi}\int_{0}^{2\pi}\frac{\cos^{2}(\theta)}
{(E + \sqrt{E^2 - Q}\sin(\theta))^4}d\theta = \frac{E(E^2 - Q)}{8\sqrt{Q}},
\end{equation}
and expressing the energy $E$ in terms of the action $J$ we find
$$A(J) = \frac{J(2J + \sqrt{Q})(J + \sqrt{Q})}{2\sqrt{Q}}.$$
Note that the expected time to random walk to infinity is
actually finite, since $A(J) \propto J^3$.
Substituting this expression for into Eq.(\ref{exit}) we obtain
for the time to pass from the bottom of the potential well where
the action $J_0 = 0$ to the state with the action $J$
\begin{equation}
J = \frac{\sqrt{Q}(e^{y} - 1)}{2 -e^y} ~~~~~~ \ y = \frac{\sigma^2 t}{2},
\end{equation}
or in the terms of the total energy
\begin{equation}\label{Egsl}
E = \frac{\sqrt{Q}e^y}{2 - e^y}.
\end{equation}
From the above it is easy to see that the expected time for the width
of the condensate to grow to infinity - the mean time for the condensate
to break up - is given by
\begin{equation}
t^{\ast} = \frac{2\ln{2}}{\sigma^2}.
\end{equation}

In Fig. \ref{fig_8} we compare the theoretical expression
Eq.(\ref{Egsl}) with numerical simulations of the full stochastic GP equation
with fluctuating scattering length in the case where $\sigma =0.04$ and
$N= 1.0$. The energy has been  numerically calculated for the localized part of
the condensate wavefunction.  We observe good agreement between
the theory and numerical simulations for $t\leq 80$.
%the theory and numerical simulations for $t\leq 30$. Again, as in the
%fluctuating trap case, the ODE solution overestimates the solution
%of the PDE.

%%%%%%%%%%%%%%  Fig.8     BEGIN    %%%%%%%%%%%%%%%%%%%%%%%%%%%%%%%
%\begin{center}
%\mbox{\hspace{-0.35in}
%\psfig{figure=fig_8.eps,height=3.8in,angle=-90.}}
%\end{center}
%  \vskip 0.2in
%{\protect\small
%   FIG.  \ref{fig_8}.
%}
%\vskip0.2in
%
%\noindent
%
%%%%%%%%%%%%%%  Fig.8     END  %%%%%%%%%%%%%%%%%%%%%%%%%%%%%%%

\section{Conclusions}
We have considered the oscillations of a 2D BEC with radial symmetry
under periodic and random modulations of the atomic scattering
length, as well as random fluctuations of trap potential.
We have calculated the position of resonances and the energy growth
using a reduced ODE for the condensate width. We have also confirmed
the analytical predictions with numerical
simulations of the 2D Gross-Pitaevskii equation.
In the resonant case the frequency of
oscillations agrees very well with the predictions of ODE, though
the amplitude shows some discrepancy at large times that is
probably due to
the approximate
nature of the variational approach. For the random modulations of
the trap potential and the atomic scattering length we have
calculated the mean exit time corresponding to the time for the
amplitude of oscillations to exceed a given value
and estimated the magnitude of the escape time for real experiments
with 2D Bose-Einstein condensates.

\section{Acknowledgements}
We acknowledge partial support by the US Civilian Research and Development
Foundation (Award ZM2-2095) and by the Uzbekistan Fundamental Science
Support Foundation (Award 15-02).

\begin{figure}
\caption{Effective potential to describe evolution of the BEC width.}
\label{fig_1}
\end{figure}

\begin{figure}
\caption{ Phase plane for the pertubation Hamiltonian at the
resonance $\Omega = 2$.}
\label{fig_2}
\end{figure}

\begin{figure}
\caption{ Oscillations of the condensate width for $\Delta\Omega/\Omega = 0.1$.
The solid line is the variational approximation, the dotted line the numerical simulations
of 2D
Gross-Pitaevskii equation (\ref{NLSE}).}
\label{fig_3}
\end{figure}

\begin{figure}
\caption{The resonant oscillations when $\Omega =2$.}
\label{fig_4}
\end{figure}

\begin{figure}
\caption{ The growth of the energy in the resonant point.}
\label{fig_5}
\end{figure}

\begin{figure}
\caption{ The oscillations of width at the second resonance $\Omega =4$}
\label{fig_6}
\end{figure}

\begin{figure}
\caption{Fluctuating trap: the growth of the energy
when $\sigma = 0.04$ and $N = 1.0$.
Solid, dot and dash lines are for the solution of PDE, ODE and
Fokker-Planck equation respectively
}
\label{fig_7}
\end{figure}

\begin{figure}
\caption{Fluctuating scattering length: the growth of the energy
when $\sigma = 0.04$ and $N = 1.0$.
Solid, dot and dash lines represent the solutions of PDE, ODE and
Fokker-Planck equation respectively
}
\label{fig_8}
\end{figure}

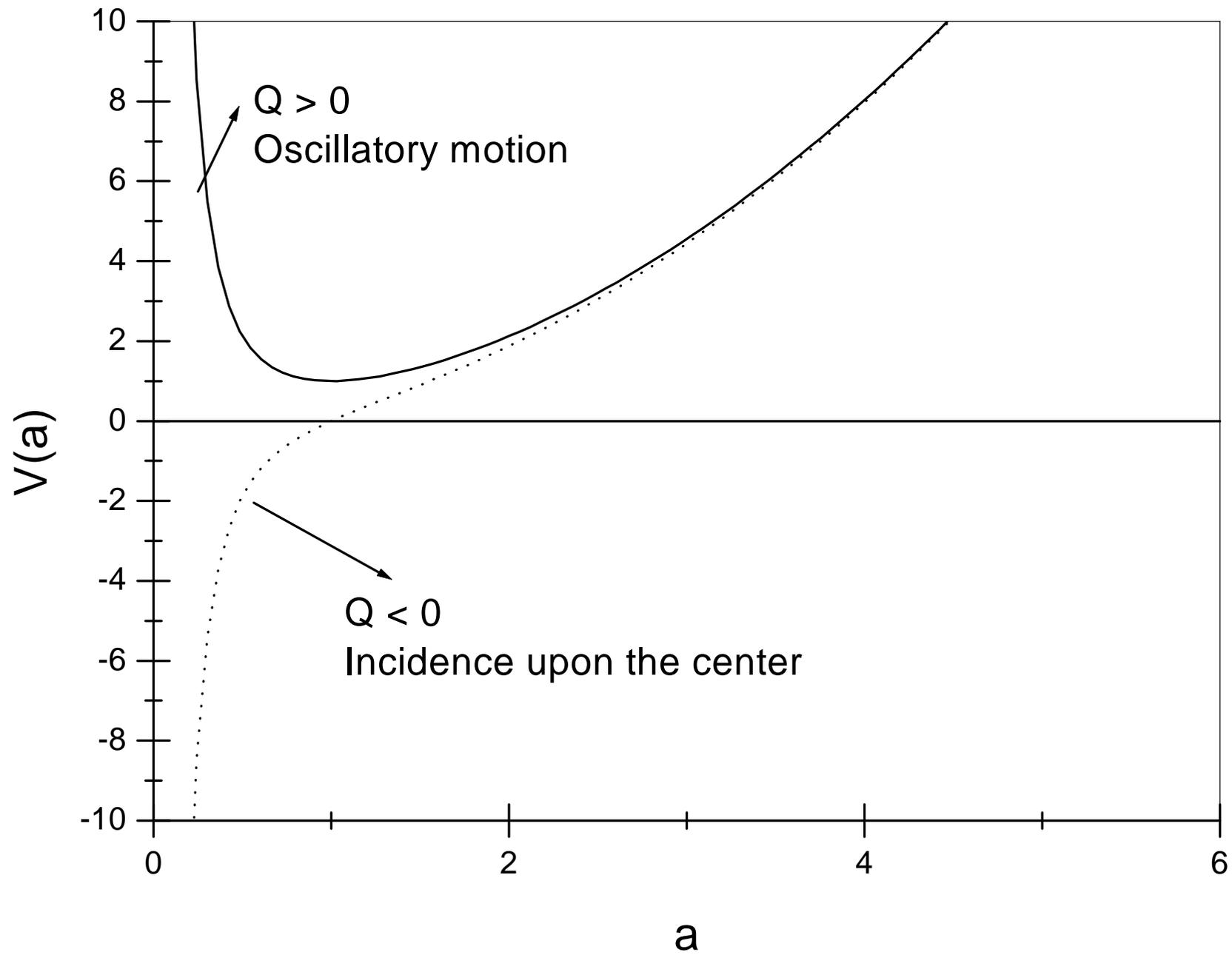
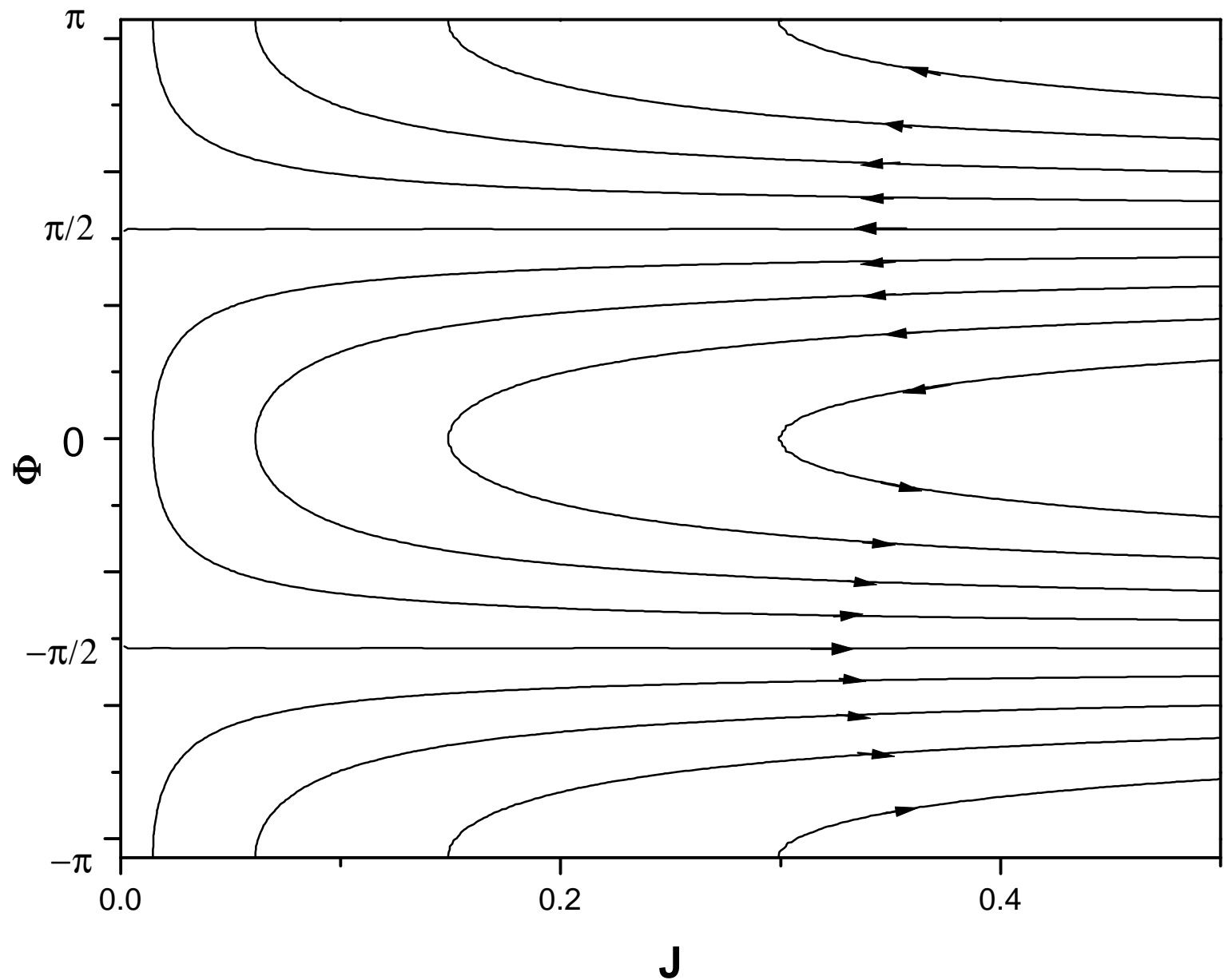
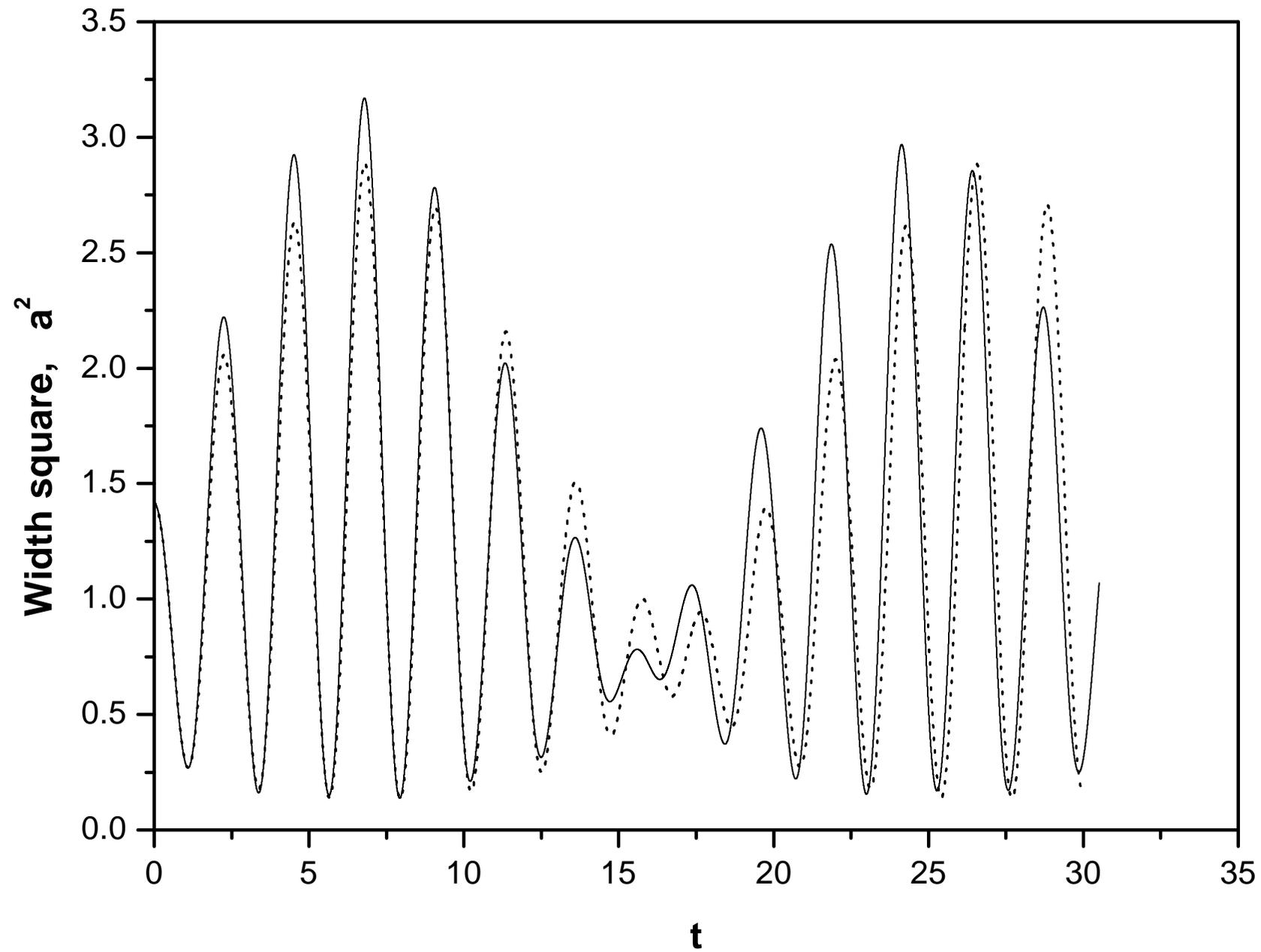
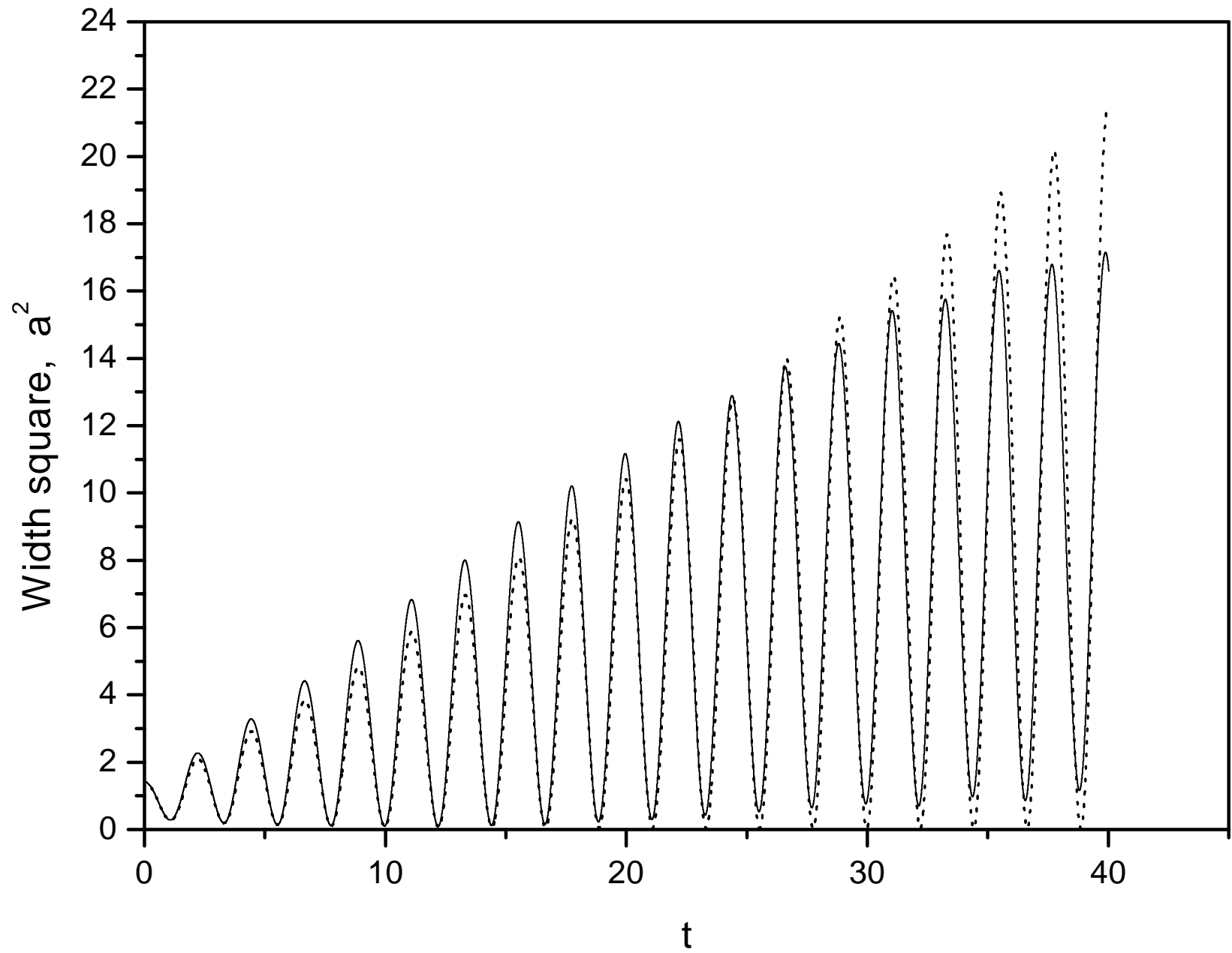
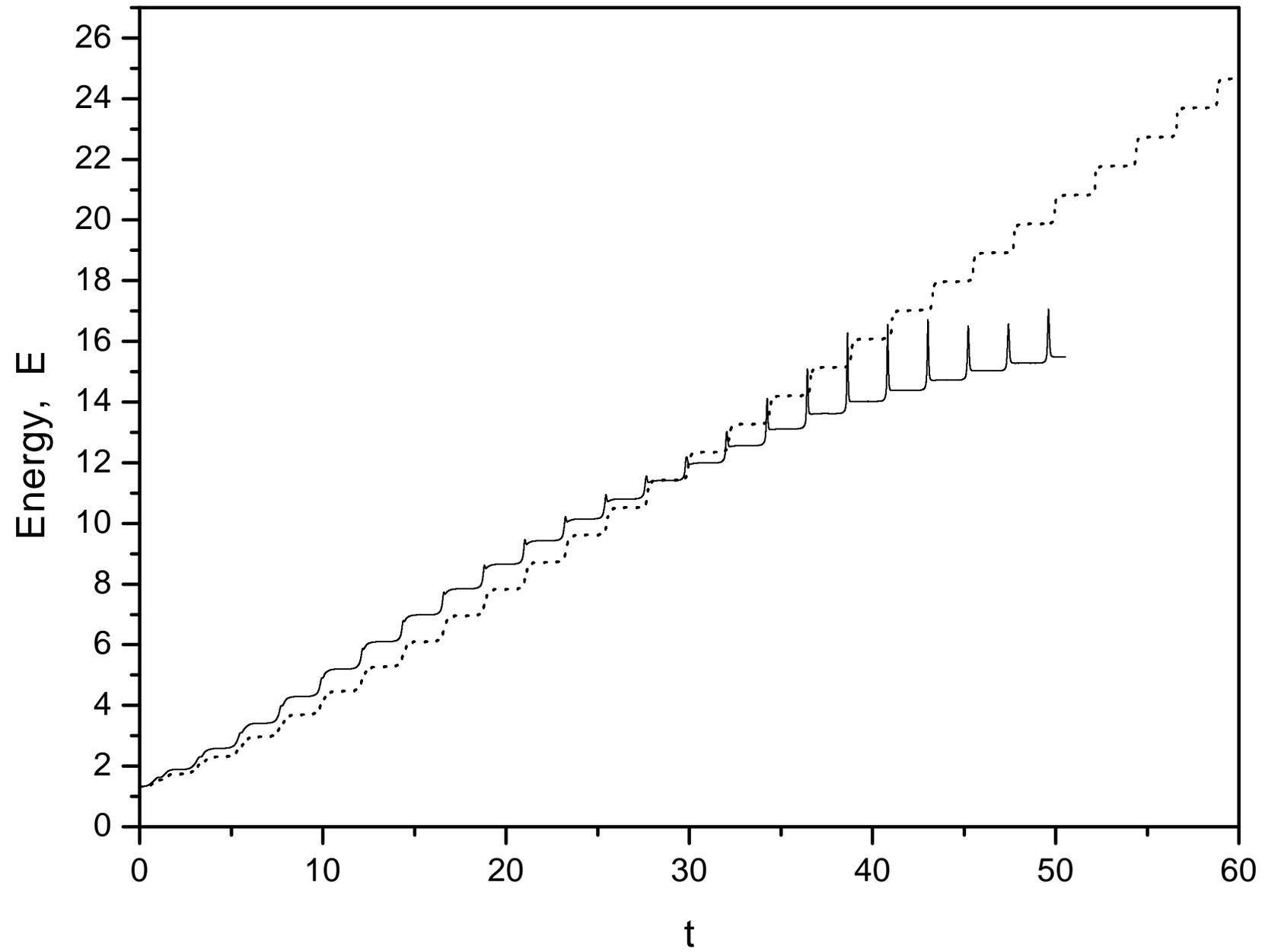
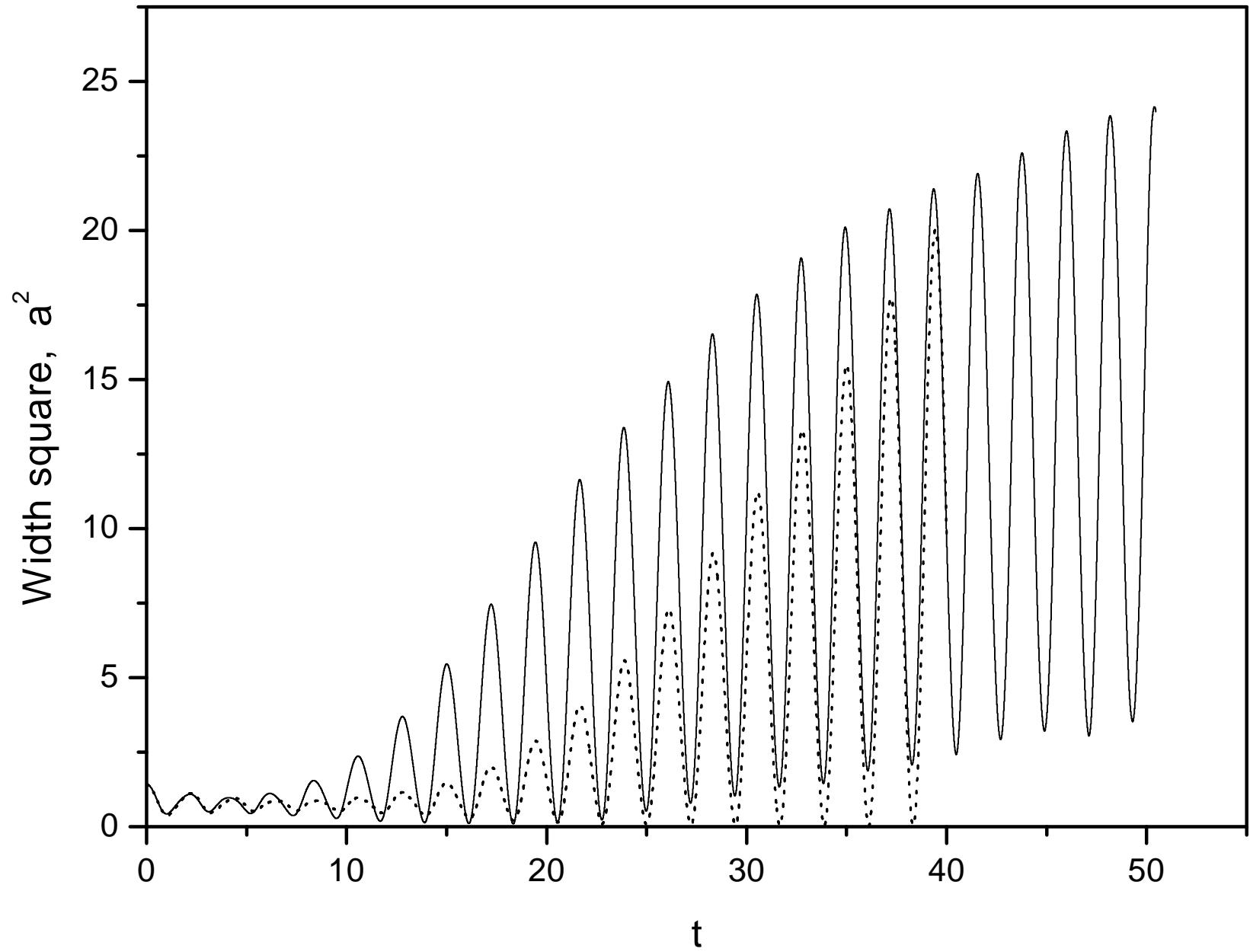
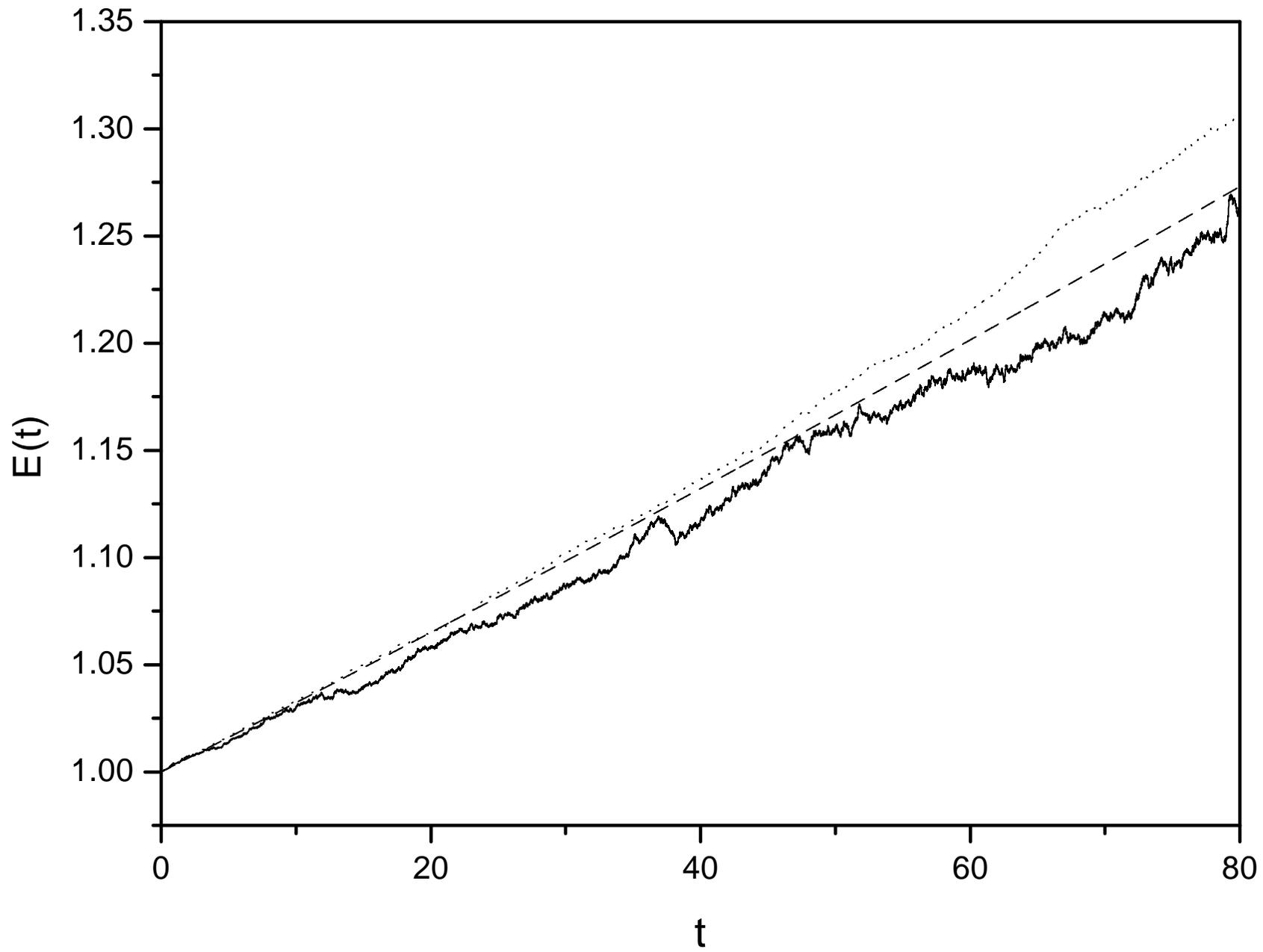
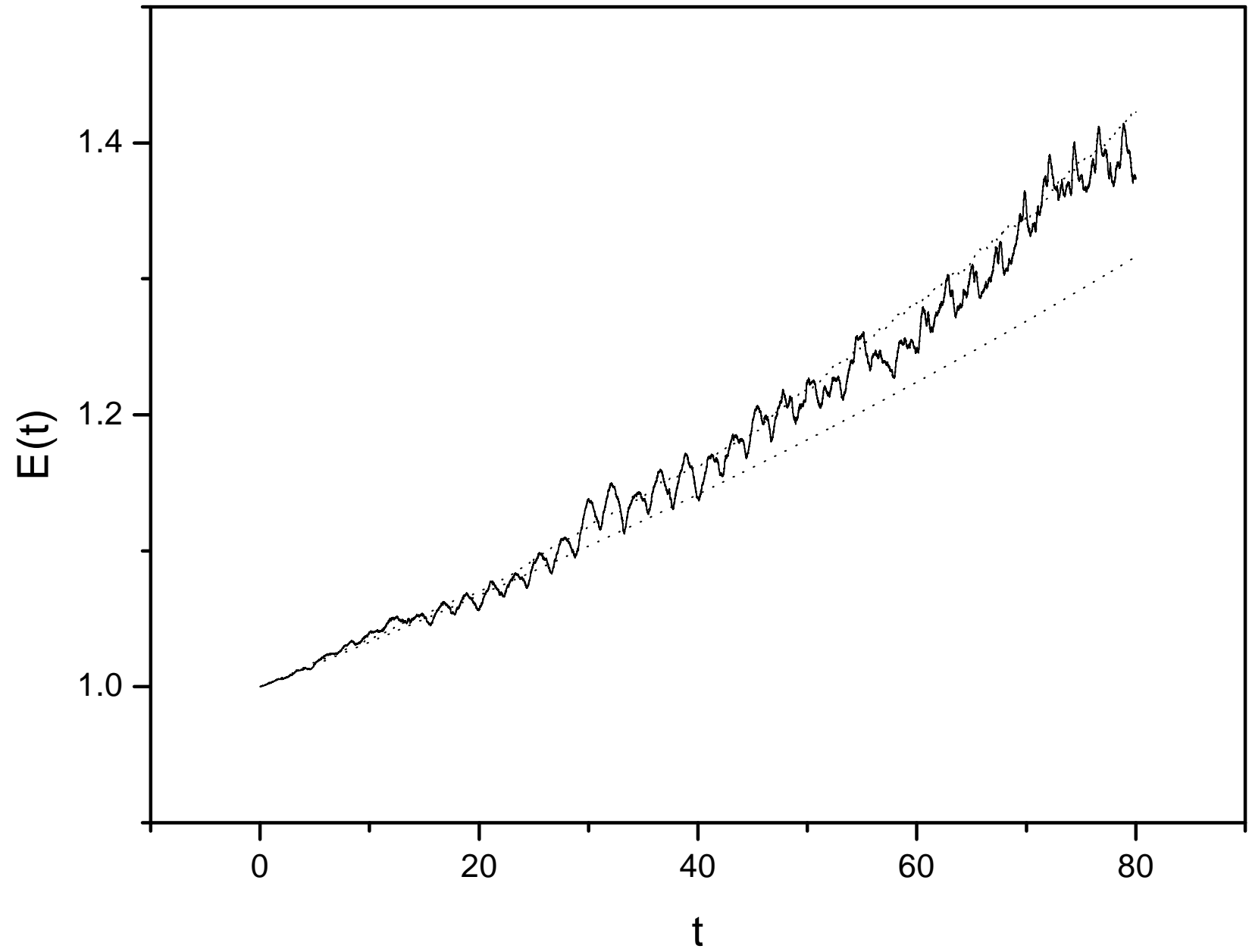
\end{document}